\documentclass[11pt]{article}

\usepackage{amsmath}
\usepackage{graphicx}
\usepackage{amsfonts}
\usepackage{amssymb}
\usepackage{epsfig}
\usepackage{color}
\usepackage{psfrag}
\usepackage{epstopdf}
\usepackage{caption}
\usepackage{subcaption}

\newcommand{\EQ}{\begin{equation}}
\newcommand{\EN}{\end{equation}}
\newcommand{\be}{\begin{equation}}
\newcommand{\ee}{\end{equation}}
\newcommand{\bea}{\begin{eqnarray}}
\newcommand{\eea}{\end{eqnarray}}

\DeclareMathOperator{\sgn}{sgn}

\begin{document} \setcounter{page}{0}
\topmargin 0pt
\oddsidemargin 5mm
\renewcommand{\thefootnote}{\arabic{footnote}}
\newpage
\setcounter{page}{0}
\topmargin 0pt
\oddsidemargin 5mm
\renewcommand{\thefootnote}{\arabic{footnote}}
\newpage

\begin{titlepage}
\begin{flushright}
\end{flushright}
\vspace{0.5cm}
\begin{center}
{\large {\bf Critical points in the $RP^{N-1}$ model
}}\\
\vspace{1.8cm}
{\large Youness Diouane$^{1,2}$, Noel Lamsen$^{1}$ and Gesualdo Delfino$^{1}$}\\
\vspace{0.5cm}
{\em $^{1}$SISSA and INFN -- Via Bonomea 265, 34136 Trieste, Italy}\\
{\em $^{2}$ICTP, Strada Costiera 11, 34151 Trieste, Italy}\\

\end{center}
\vspace{1.2cm}

\renewcommand{\thefootnote}{\arabic{footnote}}
\setcounter{footnote}{0}

\begin{abstract}
\noindent
The space of solutions of the exact renormalization group fixed point equations of the two-dimensional $RP^{N-1}$ model, which we recently obtained within the scale invariant scattering framework, is explored for continuous values of $N\geq 0$. Quasi-long-range order occurs only for $N=2$, and allows for several lines of fixed points meeting at the BKT transition point. A rich pattern of fixed points is present below $N^*=2.24421..$, while only zero temperature criticality in the $O(N(N+1)/2-1)$ universality class can occur above this value. The interpretation of an extra solution at $N=3$ requires the identitication of a path to criticality specific to this value of $N$. 
\end{abstract}
\end{titlepage}

\newpage
\tableofcontents

\section{Introduction}
The $RP^{N-1}$ lattice model differs from the $O(N)$ vector model for the additional invariance under inversion of individual spins. The consequent head-tail symmetry of the elementary degrees of freedom makes the model relevant for liquid crystals \cite{deGP}, and paradigmatic for studying the ability of an extra local symmetry to affect critical behavior. While the weak first order transition observed in numerical simulations of the three-dimensional ferromagnetic model \cite{ZMZ} is consistent with the mean field scenario \cite{deGP}, the situation is more subtle in two dimensions. Here, the effect of the fluctuations is stronger and mean field is maximally unreliable, as illustrated by the example of the three-state Potts model, whose transition becomes second order in two dimensions \cite{Wu}. In addition, since in two dimensions the continuous symmetry of the $RP^{N-1}$ model cannot break spontaneously \cite{MWHC}, finite temperature criticality could only be produced by a topological transition similar to the Berezinskii-Kosterlitz-Thouless (BKT) one \cite{BKT,Cardy_book}, and "disclination" defects \cite{Stein,Mermin} have been proposed as possible mediators of such a transition. Alternatively, zero temperature criticality can occur, analogously to what happens in the $O(N>2)$ model.  

While the presence of the local symmetry adds to the nonperturbative character of these issues, the model remained out of reach also for the traditional exact approaches in two dimensions, namely lattice integrability \cite{Baxter} and conformal field theory \cite{DfMS,BPZ}. Only in a recent Letter \cite{DDL} we showed that the renormalization group fixed points of the $RP^{N-1}$ model can be accessed exactly. This was achieved exploiting the scale invariant scattering theory \cite{paraf,sis} that in the last years allowed to obtain new results for the critical properties of pure and disordered two-dimensional systems \cite{random,DT1,DL_ON_jhep,DL_ON,DL_vector_scalar,DL_softening}. The novelty of the method is that conformal invariance, which in two dimensions has infinitely many generators, is implemented in the basis of particle excitations, and this yields exact equations for the renormalization group fixed points. In this paper we explore the space of solutions of the fixed point equations of the $RP^{N-1}$ model for $N\geq 0$. The analysis is performed for continuous values of $N$, and allows to distinguish the range $N<N^*=2.24421..$ characterized by a rich pattern of fixed points from the range $N>N^*$ in which the equations generically possess a single solution. Quasi-long-range order is found only at $N=2$, but turns out to allow for several lines of fixed points meeting at the BKT transition point. For $N\geq 3$ we confirm and further discuss the results anticipated in \cite{DDL}. 

The paper is organized as follows. In the next section we review the scale invariant scattering formalism and illustrate its application on the example of the $O(M)$ model, which is relevant for the continuation of the analysis. Section~3 is devoted to the $RP^{N-1}$ fixed point equations, whose solutions are then given in section~4. Section~5 contains additional remarks, while more technical aspects are presented in the appendices.

\section{Scale invariant scattering}
\subsection{Generalities}
Before turning to the $RP^{N-1}$ model, in this section we recall the generalities of the scale invariant scattering theory \cite{paraf,sis}, as well as its application to the $O(M)$ vector model. Scale invariant scattering exploits the fact that a two-dimensional statistical system at criticality is described in the continuum limit by a Euclidean field theory, and that the latter is the continuation to imaginary time of a conformally invariant quantum field theory with one space and one time dimension. Such a quantum theory possesses massless particles describing the fluctuation modes of the system. The fact that conformal symmetry in two dimensions possesses infinitely many generators implies an infinite number of conservation laws for the scattering processes of the particles, with the result that scattering is completely elastic (i.e. initial and final states are kinematically identical). Moreover, the fact that the center of mass energy is the only relativistic invariant of two-particle scattering and is dimensionful leaves an energy-independent scattering amplitude for the critical system as a consequence of scale invariance. 

\begin{figure}
     \begin{subfigure}[b]{.5\textwidth}
         \centering
         \includegraphics[width=.4\linewidth]{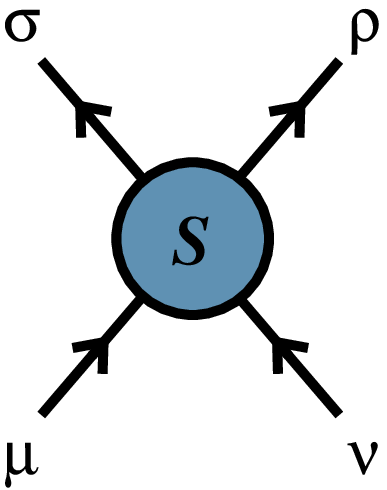}
     \end{subfigure}
     \hfill
     \begin{subfigure}[b]{.5\textwidth}
         \centering
         \includegraphics[width=.3\linewidth]{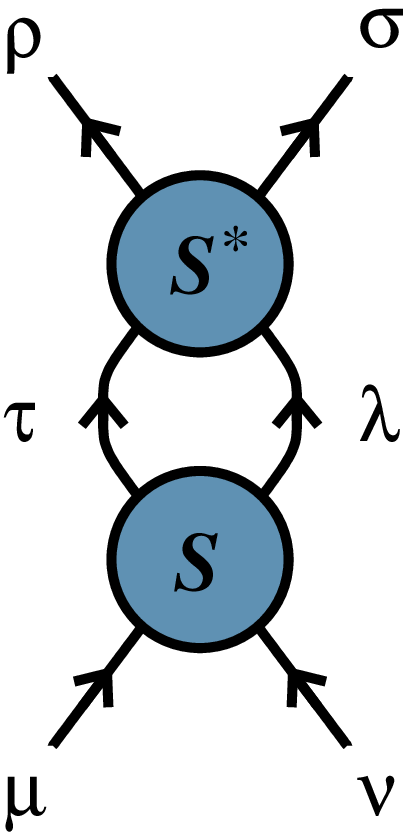}
     \end{subfigure}
\caption{Scattering amplitude $S_{\mu\nu}^{\rho\sigma}$ (left) and product of amplitudes appearing in the unitarity equations (\ref{unitarity}) (right).  
}
\label{scattering}
\end{figure}

Such special features of two-dimensional criticality are responsible for a  substantial simplification of the unitarity and crossing equations that generally apply to relativistic scattering \cite{ELOP,fpu}. Indeed, if we denote by $\mu=1,2,\ldots,k$ the particle species\footnote{In this paper we can limit our discussion to self-conjugated particles.}, by $\mathbb{S}$ the scattering operator, and by $S_{\mu\nu}^{\rho\sigma}=\langle\rho\sigma|\mathbb{S}|\mu\nu\rangle$ the amplitude for a scattering process with particles $\mu$ and $\nu$ in the initial state and particles $\rho$ and $\sigma$ in the final state (figure~\ref{scattering}), we have \cite{paraf}
\EQ
S_{\mu\nu}^{\rho\sigma}=\left[S_{\mu\sigma}^{\rho\nu}\right]^*
\label{crossing}
\EN
for crossing and
\EQ
\sum_{\lambda,\tau}S_{\mu\nu}^{\lambda\tau}\left[S_{\lambda\tau}^{\rho\sigma}\right]^*= \delta_{\mu\rho}\delta_{\nu\sigma}\,
\label{unitarity}
\EN
for unitarity. The relations
\EQ
S_{\mu\nu}^{\rho\sigma}=S_{\rho\sigma}^{\mu\nu}=S_{\nu\mu}^{\sigma\rho}\,
\label{TS}
\EN
also hold and express invariance of the amplitudes under time reversal and spatial inversion.

\subsection{$O(M)$ model}
\begin{figure}
\begin{center}
\includegraphics[width=8cm]{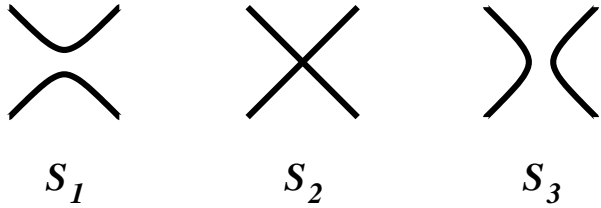}
\caption{Scattering amplitudes entering (\ref{ON}). Time runs upwards.
}
\label{vector_ampl}
\end{center} 
\end{figure}

Consider now the $O(M)$ model \cite{paraf,DL_ON}. It is defined on the lattice by the Hamiltonian
\EQ
{\cal H}_1=-J\sum_{\langle i,j\rangle}{\bf s}_i\cdot{\bf s}_j\,,
\label{lattice_ON}
\EN
where $J$ is the coupling, ${\bf s}_i$ is a $M$-component unit vector located at site $i$, and the sum is taken over nearest neighboring sites. In the scattering description the order parameter variable ${\bf s}_i$ corresponds to a vector multiplet of particles labeled by an index $a=1,2,\ldots,M$. The scattering of a particle $a$ with a particle $b$ involves a tensorial structure that has to be preserved by the scattering. The $O(M)$ scattering matrix is then written as
\begin{equation}
S_{ab}^{cd}=S_1\,\delta_{ab}\delta_{cd}+S_2\,\delta_{ac}\delta_{bd}+S_3\,\delta_{ad}\delta_{bc}\,,
\label{ON}
\end{equation}
where the amplitudes $S_1$, $S_2$ and $S_3$ correspond to annihilation, transmission and reflection, respectively; they are depicted in figure~\ref{vector_ampl}. The crossing symmetry equations (\ref{crossing}) translate into the relations
\bea
S_1=S_3^{*} &\equiv &  \rho_{1}\,e^{i\phi}, 
\label{cr1}\\
S_2 = S_2^* &\equiv & \rho_2,
\label{cr2}
\eea 
where we introduced the parametrizations in terms of the variables $\rho_2$ and $\phi$ real, and $\rho_1\geq 0$. This in turn allows us to write the unitarity equations (\ref{unitarity}) in the form
\bea
&& \rho_1^2+\rho_2^2=1\,,  \label{u1}\\
&& \rho_1 \rho_2 \cos\phi=0\,,\label{u2}  \\
&& M \rho_1^2 + 2\rho_1\rho_2 \cos\phi +2\rho_1^2\cos2\phi=0\,. \label{u3} 
\eea

\noindent
The solutions of equations (\ref{u1})-(\ref{u3}) \cite{paraf,DL_ON}, which are listed in table~\ref{solutions}, yield the renormalization group fixed points with $O(M)$ symmetry. Referring the reader to \cite{DL_ON} for the detailed discussion of the solutions, here we recall some main points useful for the subsequent sections. Starting from the two solutions II$_\pm$, we notice that they are characterized by $S_2=0$, namely by absence of intersection of particle trajectories (see figure \ref{vector_ampl}); in addition, they are defined in the range $M\in[-2,2]$, and meet at $M=2$. These properties identify them as the critical lines of the dilute and dense regimes of the loop gas model, for which the mapping of the partition function onto that of the $O(M)$ model is well known \cite{Cardy_book,Nienhuis}. In particular, the loop formulation realizes on the lattice the continuation to noninteger values of $M$ that equations (\ref{u1})-(\ref{u3})  provide directly in the continuum; self-avoiding walks are obtained in the limit $M\to 0$ \cite{DeGennes}. Nonintersection of loop paths corresponds to nonintersection of particle trajectories, as originally pointed out in \cite{Zamo_SAW} for the off-critical case.

\begin{table}
\begin{center}
\begin{tabular}{l|c|c|c|c}
\hline 
Solution & $M$ & $\rho_1$ & $\rho_2$ & $\cos\phi$ \\ 
\hline \hline
I$_{\pm}$ & $(-\infty,\infty)$ & $0$ & $\pm 1$ & -  \\ 
II$_{\pm}$ & $[-2, 2]$ & $1$ & $0$ & $\pm\frac{1}{2}\sqrt{2-M}$  \\ 
III$_{\pm}$ & $2$ & $[0,1]$ & $\pm \sqrt{1- \rho_1^2}$ & $0$ \\[0.7em] 
\hline 
\end{tabular} 
\caption{Solutions of equations (\ref{u1})-(\ref{u3}), yielding the renormalization group fixed points with $O(M)$ symmetry.
}
\label{solutions}
\end{center}
\end{table}


The solutions III$_\pm$, which are defined only for $M=2$, are characterized by the fact of containing $\rho_1$ as a free parameter. Hence, they are immediately identified with the line of fixed points responsible for the BKT transition in the $O(2)$ ferromagnet \cite{BKT,Cardy_book}. The meeting point $\rho_1=1$ of III$_+$ and III$_-$ corresponds to the BKT transition point, where the field driving the transition is marginal in the renormalizaion group sense \cite{paraf,DL_ON}. This field is instead irrelevant along III$_+$, so that this is the BKT phase in which correlations decay algebraically (quasi-long-range order). 

The remaining solutions I$_\pm$ only allow for transmission. Since scattering on a line involves position exchange and mixes interaction and statistics, these solutions correspond to noninteracting bosons for $S_2=1$, and noninteracting fermions for $S_2=-1$. The bosonic solution I$_+$ describes the $T=0$ critical point of the nonlinear sigma model with reduced Hamiltonian
\EQ
{\cal H}_{SM}=\frac{1}{T}\int d^2x\left(\nabla{\bf s}\right)^2, \hspace{1cm}{\bf s}^2=1\,,
\label{sigma}
\EN
where ${\bf s}(x)$ is the counterpart in the continuum of the lattice variable ${\bf s}_i$ and $T$ is the temperature. When $M>2$ the sigma model exhibits exponentially diverging correlation length and vanishing interaction for $T\to 0$ (asymptotic freedom) \cite{Cardy_book,Zinn}; it describes the continuum limit of the $O(M)$ model in this range of $M$. Consistency requires that the zero temperature endpoint $\rho_1=0$ of the BKT phase III$_+$ coincides with I$_+$, a property that is indeed exhibited by the solutions of table~\ref{solutions}. On the other hand, the solution I$_-$ is not relevant for the critical behavior of the vector model (\ref{lattice_ON}), since it yields a realization of the symmetry in terms of $M$ fermions.

\section{Fixed point equations of the $RP^{N-1}$ model}
The $RP^{N-1}$ lattice model is  defined by the Hamiltonian
\EQ
{\cal H}_2=-J\sum_{\langle i,j\rangle}({\bf s}_i\cdot{\bf s}_j)^2\,,
\label{lattice}
\EN
where ${\bf s}_i$ is a $N$-component unit vector located at site $i$. The difference with respect to the Hamiltonian (\ref{lattice_ON}) is the square in the r.h.s., which makes (\ref{lattice}) invariant under any local reversal ${\bf s}_i\to -{\bf s}_i$, thus ensuring head-tail symmetry. This means that ${\bf s}_i$ effectively takes values on the unit hypersphere with opposite points identified, namely in the real projective space the model is named after. The global and local symmetries of the model are represented through an order parameter variable quadratic in the vector components $s_i^a$, namely by the symmetric tensor \cite{deGP}
\EQ
Q^{ab}_i=s^a_i s^b_i-\frac{1}{N}\delta_{ab}\,.
\label{op}
\EN
While $\sum_a s^a_i s^a_i=1$ excludes the presence of an invariant linear in the order parameter components, $\textrm{Tr}\,Q^{ab}_i=0$ ensures that, upon diagonalization in generic dimension, the order parameter $\langle Q^{ab}_i\rangle$ vanishes in the isotropic phase. We denote by $\langle\cdots\rangle$ the average over configurations weighted by $e^{-{\cal H}_2/T}$. 

\begin{figure}
\begin{center}
\includegraphics[width=15cm]{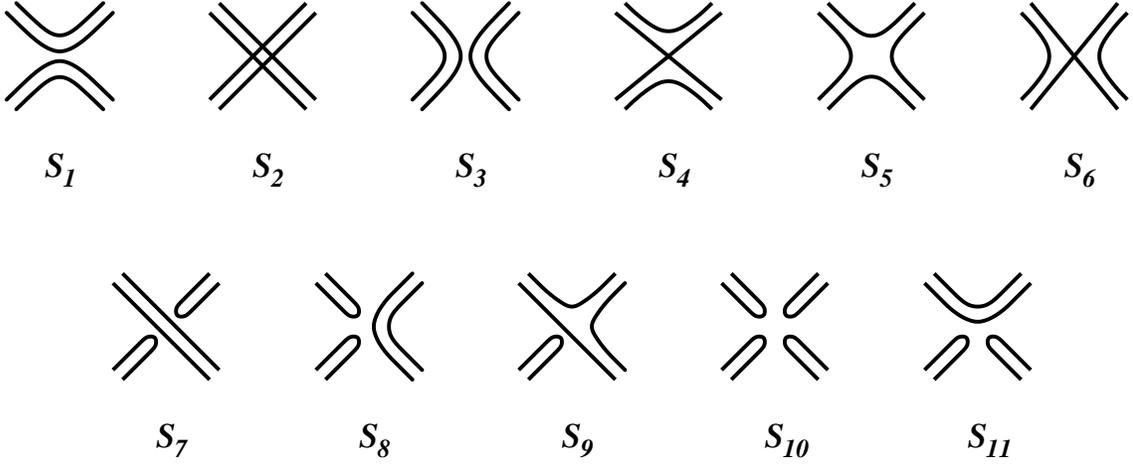}
\caption{Scattering amplitudes appearing in (\ref{S_tensor}). Time runs upwards.
}
\label{tensor_ampl}
\end{center} 
\end{figure}

The steps through which we implement scale invariant scattering for the two-dimensional $RP^{N-1}$ model at criticality parallel those seen in the previous section for the vector model. In the continuum limit, the order parameter field is now the symmetric tensor $Q_{ab}(x)$, which creates particles that we label by $\mu=ab$, with $a$ and $b$ going from 1 to $N$. The scattering processes corresponding to these particles are those shown in figure~\ref{tensor_ampl}. Taking into account also the relations (\ref{TS}), the scattering matrix is expressed in terms of the amplitudes $S_1,\ldots,S_{11}$ as
\begin{equation}
\begin{split}
S_{ab, cd}^{ef,gh} &= S_1\,\delta_{(ab),(cd)}^{(2)}\delta_{(ef),(gh)}^{(2)} + S_2\,\delta_{(ab), (ef)}^{(2)} \delta_{(cd),(gh)}^{(2)} + S_3\,\delta_{(ab),(gh)}^{(2)}\delta_{(cd),(ef)}^{(2)}\\
& + S_4\,\delta_{(ab)(gh),(cd)(ef)}^{(4)}  + S_5\,\delta_{(ab)(ef),(cd)(gh)}^{(4)} + S_6\,\delta_{(ab)(cd),(ef)(gh)}^{(4)}\\
& +S_7\left[\delta_{ab}\delta_{ef}\delta_{(cd),(gh)}^{(2)}+\delta_{cd}\delta_{gh}\delta_{(ab),(ef)}^{(2)}\right]  +S_8\left[\delta_{ab}\delta_{gh}\delta_{(cd),(ef)}^{(2)}+\delta_{cd}\delta_{ef}\delta_{(ab),(gh)}^{(2)}\right]\\
& +S_9\left[\delta_{ab}\delta_{(cd),(ef),(gh)}^{(3)}+\delta_{cd}\delta_{(ab),(ef),(gh)}^{(3)}+\delta_{ef}\delta_{(cd),(ab),(gh)}^{(3)}+\delta_{gh}\delta_{(cd),(ef),(ab)}^{(3)}\right]\\
& + S_{10}\,\delta_{ab}\delta_{cd}\delta_{ef}\delta_{gh}+S_{11}\left[\delta_{ab}\delta_{cd}\delta_{(ef),(gh)}^2+\delta_{ef}\delta_{gh}\delta_{(ab),(cd)}^{(2)}\right],
\end{split}
\label{S_tensor}
\end{equation}
where we introduced the notations
\begin{align}
\delta^{(2)}_{(ab),(cd)} & \equiv(\delta_{ac} \delta_{bd} + \delta_{ad} \delta_{bc})/2\,,
\label{delta2}\\
\label{delta3}\delta^{(3)}_{(ab),(cd),(ef)} &\equiv(\delta_{af}\delta_{bd}\delta_{ce} + \delta_{ad}\delta_{bf}\delta_{ce}+\delta_{ae}\delta_{bd}\delta_{cf} + \delta_{ad}\delta_{be}\delta_{cf} \nonumber\\ 
&+ \delta_{af}\delta_{bc}\delta_{de} + \delta _{ac} \delta_{bf} \delta_{de}+\delta _{ae} \delta _{bc} \delta _{df}+\delta _{ac} \delta _{be} \delta_{df})/8\,,\\
\label{delta4}\delta^{(4)}_{(ab)(cd),(ef)(gh)} &\equiv (\delta _{ah} \delta _{bf} \delta _{cg} \delta _{de}+\delta _{af} \delta _{bh}
   \delta _{cg} \delta _{de}+\delta _{ag} \delta _{bf} \delta _{ch} \delta_{de}+\delta _{af} \delta _{bg} \delta _{ch} \delta _{de}\nonumber \\
   &+\delta _{ah} \delta_{be} \delta _{c,g} \delta _{df}+\delta _{a,e} \delta _{bh} \delta _{cg} \delta_{df}+\delta_{ag} \delta _{be} \delta _{ch} \delta _{df}+\delta _{ae} \delta_{bg} \delta _{ch} \delta _{df} \nonumber\\ 
   &+\delta _{ah} \delta _{bf} \delta _{ce} \delta_{dg}+\delta _{af} \delta _{bh} \delta _{ce} \delta_{dg}+\delta_{ah} \delta_{be} \delta _{cf} \delta _{dg}+\delta _{ae} \delta _{bh} \delta _{cf} \delta_{dg}\nonumber \\ 
   &+\delta _{ag} \delta _{bf} \delta _{ce} \delta _{dh}+\delta _{af} \delta_{bg} \delta _{ce} \delta _{dh}+\delta _{ag} \delta _{be} \delta _{cf} \delta_{dh}+\delta _{ae} \delta _{bg} \delta _{cf} \delta _{dh})/4\,
\end{align}
to take into account the different ways of contracting the particle indices for a given process in figure~\ref{tensor_ampl}. The fact that the indices of a particle $aa$ can annihilate each other gives rise to the amplitudes $S_{i\geq 7}$.

Since the amplitudes $S_{i\leq 3}$ satisfy the crossing equations (\ref{cr1}) and (\ref{cr2}), we keep for them the same parametrization in terms of $\rho_1$, $\rho_2$ and $\phi$. On the other hand, the crossing relations for the remaining amplitudes lead to the parametrizations
\bea
S_4=S_6^* &\equiv & \rho_4 e^{i\theta}\,,\\
S_5=S_5^* &\equiv & \rho_5\,,\\
S_7=S_7^* &\equiv & \rho_7\,,\\
S_8=S_{11}^* &\equiv & \rho_8 e^{i\psi}\,,\\
S_9=S_9^* &\equiv & \rho_9\,,\\
S_{10}=S_{10}^* &\equiv & \rho_{10}\,.
\eea
The unitarity equations (\ref{unitarity}) now take the explicit form
\EQ
\label{unitaritytensor} \sum\limits_{e,f,g,h=1}^NS_{ab,cd}^{ef,gh}\left[S_{ef,gh}^{a'b',c'd'}\right]^*=\delta_{(ab),(a'b')}^{(2)}\delta_{(cd),(c'd')}^{(2)}\,,
\EN
which takes into account the present way of indexing the particles. The different possible choices of external indices (see table~\ref{indices}) then yield the equations
\begin{align}
1 &= \rho_1^2 + \rho_2^2 + 4\rho_4^2 \label{eq:u1symmat},\\
0 &= 2 \rho_1\rho_2\cos\phi + 4\rho_4^2 \label{eq:u2symmat},\\
0 &= (M_N+1)\rho_1^2 + 2\rho_1^2\cos 2\phi + 2\rho_1\rho_2\cos\phi + 4\rho_1\rho_4\cos(\phi + \theta) + 4( \rho_4^2 + \rho_5^2 + 2\rho_4\rho_5\cos\theta  )  \nonumber\\
&  \quad + 4(N+1) (\rho_1\rho_4\cos(\phi - \theta) + \rho_1\rho_5\cos\phi) + 2N\rho_1\rho_8\cos(\phi + \psi) + 8\rho_4\rho_8\cos(\theta - \psi) \nonumber\\
&  \quad + 8\rho_4\rho_8\cos(\theta + \psi) + 8\rho_5\rho_8\cos\psi  + N^2\rho_8^2 + 4\rho_1\rho_9\cos\phi + 4N\rho_8\rho_9\cos\psi + 2\rho_9^2 \label{eq:u3symmat}, \\
0 &= 2\rho_2\rho_5 + 2\rho_1\rho_4\cos(\phi + \theta) + 2\rho_4^2\cos 2\theta + 2(N+3)\rho_4\rho_5\cos\theta 
+ 4\rho_4\rho_9\cos\theta  \nonumber\\
& \qquad + 2 \rho_5\rho_9 + \tfrac{1}{4}N\rho_9^2 \label{eq:u4symmat}, \\
0 &= 2\rho_1\rho_5\cos\phi + 2\rho_2\rho_4\cos\theta  +  2\rho_4^2\cos 2\theta + 2\rho_4\rho_5\cos\theta + (N+2)(\rho_4^2 + \rho_5^2) \nonumber \\
& \quad + 4\rho_4\rho_9\cos\theta + 2\rho_5 \rho_9 + \tfrac{1}{4}N\rho_9^2 \label{eq:u5symmat},\\
0 &= 2 \rho_1 \rho_4 \cos(\phi - \theta) + 2\rho_2 \rho_4 \cos\theta + 2\rho_4^2 \label{eq:u6symmat}, \\
0 &= 2\rho_1 \rho_7 \cos\phi + 2 \rho_2 \rho_8 \cos\psi +  2 N \rho_7 \rho_8 \cos\psi+ 2 \rho_4 \rho_9 \cos\theta + 2\rho_7 \rho_9 + 2\rho_8 \rho_9 \cos\psi \nonumber\\ 
& \qquad+ \tfrac{1}{4}(N+2) \rho_9^2 \label{eq:u7symmat}, \\
0 &= 2 \rho_1 \rho_8 \cos(\phi + \psi) + 2 \rho_2 \rho_7 + N (\rho_7^2 + \rho_8^2)
+ 2\rho_4 \rho_9 \cos\theta + 2 \rho_7 \rho_9 + 2 \rho_8 \rho_9 \cos\psi\nonumber\\
& \qquad + \tfrac{1}{4} (N+2) \rho_9^2\label{eq:u8symmat}, \\
0&= 4 \rho _8 \rho _4 e^{i \psi } \cos \theta +2 e^{-2 i \theta } \rho _4^2+2 e^{-i \theta } \rho _5 \rho _4+4 \rho _7 \rho _4 \cos \theta +2 \rho _9 \rho _4 \cos \theta +\frac{1}{2} e^{-i \theta } N \rho _9 \rho _4\nonumber\\
   & \qquad+\frac{1}{2} N \rho _8 \rho _9 e^{i \psi }+\left(\frac{N}{2}+1\right) \rho _5 \rho
   _9+\frac{1}{2} N \rho _7 \rho _9+2 \rho _5 \rho _8 e^{i \psi }+\rho _1 \rho _9 \cos \phi +\rho _9^2+2 \rho _5 \rho _7\nonumber\\
& \qquad+\rho _2 \rho _9\label{eq:u9symmat}, 
\end{align}
\begin{align}
0 &= 4 \rho _4 \rho _8 \cos (\theta -\psi )+ \left(M_N+3\right) \rho _8^2+N^2 \rho_{10}^2+2 (N+1) \rho _8 \rho _9 \cos \psi + 6 N \rho _8 \rho _{10} \cos  \psi  \nonumber\\ 
&  \quad + 4N\rho_7\rho_{10} + 8 \rho_7 \rho_8 \cos\psi + 4\rho_8^2 \cos  2 \psi  +2 \rho _1 \rho _{10} \cos \phi  +2 \rho _7^2+\rho _9^2+2 \rho _2 \rho _{10} + 4 \rho _9 \rho _{10} \label{eq:u10symmat}, \\
0&= 4\rho _1 \rho _4e^{-i (\theta +\phi )}  +4 \rho _4 \rho _9e^{-i \theta } +16  \rho _4 \rho _{10}\cos \theta + 2(M_N+1) \rho _1 \rho _8e^{-i (\psi+\phi )}+4\rho _1 \rho _8 e^{-i (\phi -\psi )}\nonumber\\
   & \qquad+4 \rho _1 \rho _8\cos (\phi -\psi )+2 N^2 \rho _8 \rho _{10}  e^{i \psi }+4   \left(2 \cos \theta+e^{-i \theta } N\right)\rho _4 \rho _8e^{-i \psi }+N  \left(2+4 e^{2 i \psi }\right)\rho _8^2\nonumber\\
& \qquad+4 N  \rho _7 \rho _8e^{i \psi }+4 (N+1) \rho _5 \rho _8 e^{-i \psi }+2 N \rho _1\rho _{10} e^{-i \phi }+2 (N+1) \rho _1 \rho _9 e^{-i \phi }+4 N \rho _9 \rho _{10}\nonumber\\
& \qquad+4  \left(2\cos\psi+ e^{i \psi }\right)\rho _8 \rho _9+4  \rho _2 \rho _8\cos \psi +4 \rho _1 \rho _7 e^{-i \phi }+4 \rho _5 \rho _9+4 \rho _7 \rho _9+8 \rho _5 \rho _{10}\label{eq:u11symmat}\,
\end{align}
where
\EQ
M_N\equiv\frac{1}{2}N(N+1)-1\,
\label{M_N}
\EN
coincides with the number of independent components of the order parameter variable (\ref{op}). In table~\ref{indices} different latin letters correspond to different values from 1 to $N$; we checked that no new constraints arise from different choices. 

\begin{table}[t!]
\centering
\begin{tabular}{|c|c|c|c|c|}
\hline
Equation & $\mu$ & $\nu$ & $\rho$ & $\sigma$\\
\hline
\hline
\eqref{eq:u1symmat} & $ab$ & $cd$ & $ab$ & $cd$\\
\hline
\eqref{eq:u2symmat} & $ab$ & $cd$ & $cd$ & $ab$\\
\hline
\eqref{eq:u3symmat} & $ab$ & $ba$ & $cd$ & $dc$\\
\hline
\eqref{eq:u4symmat} & $ab$ & $bc$ & $cd$ & $da$\\
\hline
\eqref{eq:u5symmat} & $ab$ & $bc$ & $ad$ & $dc$\\
\hline
\eqref{eq:u6symmat} & $ab$ & $cd$ & $ac$ & $bd$\\
\hline
\eqref{eq:u7symmat} & $aa$ & $bc$ & $bc$ & $dd$\\
\hline
\eqref{eq:u8symmat} & $aa$ & $bc$ & $dd$ & $bc$\\
\hline
\eqref{eq:u9symmat} & $aa$ & $bc$ & $bd$ & $dc$\\
\hline
\eqref{eq:u10symmat} & $aa$ & $bb$ & $cc$ & $dd$\\
\hline
\eqref{eq:u11symmat} & $aa$ & $bb$ & $cd$ & $dc$\\
\hline
\end{tabular}
\caption{External indices used in \eqref{unitarity} to obtain the equations \eqref{eq:u1symmat}-\eqref{eq:u11symmat}.}
\label{indices}
\end{table}

At this stage we did not yet take into account the fact that the field $Q_{ab}(x)$ that creates the particles is traceless. We do this now  defining ${\cal T}=\sum_a aa$ and requiring
\EQ
\mathbb{S}|(ab){\cal T}\rangle=S_0|(ab){\cal T}\rangle\,,\hspace{1cm}S_0=\pm 1
\label{decoupling}
\EN
for any particle state $|(ab)\rangle=|ab\rangle+|ba\rangle$. In other words, we require that the trace mode ${\cal T}$ is a noninteracting\footnote{The sign factor $S_0$ takes into account that the trace mode can decouple as a free boson or a free fermion.} (and then decoupled) particle that can be discarded, thus yielding the desired sector with $\textrm{Tr}\,Q_{ab}=0$. The condition~(\ref{decoupling}) gives the relations
\begin{align}
&& S_2 + S_9 + NS_7 -S_0=
S_1 + S_9 + NS_{11}= S_3 + S_9 + NS_{8}= \nonumber\\
&& 4(S_4 + S_5 + S_6) + NS_9=
S_7 + S_8 + S_{11} + NS_{10}= 0\,,\label{decouplingconditions}
\end{align}
which we use to express the amplitudes $S_{i\geq 7}$ in terms of $S_{i\leq 6}$, namely
\begin{align}
\rho_7 &= -\frac{1}{N}(\rho_2 -S_0) + \frac{4}{N^2}(2\rho_4 \cos \theta + \rho_5) \label{sub1}, \\
\rho_8\cos \psi &= -\frac{1}{N}\rho_1 \cos \phi + \frac{4}{N^2}(2\rho_4 \cos \theta + \rho_5) \label{sub2}, \\
\rho_8 \sin \psi &= \frac{1}{N}\rho_1 \sin \phi \label{sub3}, \\
\rho_9 &= -\frac{4}{N}(2\rho_4 \cos \theta + \rho_5) \label{sub4}, \\
\rho_{10} &= \frac{1}{N^2} \Big ( 2 \rho_1 \cos \phi + \rho_2 -S_0 -\frac{12}{N}(2 \rho_4 \cos \theta + \rho_5) \Big ). \label{sub5}
\end{align}
Upon substitution of these expressions in \eqref{eq:u1symmat}-\eqref{eq:u11symmat}, the imaginary parts of \eqref{eq:u9symmat} and \eqref{eq:u11symmat} vanish, while the real parts as well as the equations \eqref{eq:u7symmat}, \eqref{eq:u8symmat} and \eqref{eq:u10symmat} become linear combinations of \eqref{eq:u1symmat}-\eqref{eq:u6symmat}. The latter are the only remaining equations and can be written in the form
\begin{align}
1&=\rho_1^2+\rho_2^2+4\rho_4^2\,, \label{uni1}\\
0&=2\rho_1\rho_2\cos\phi+4\rho_4^2\,,\label{uni2}\\
0&=M_N\rho_1^2+2\rho_1^2\cos2\phi+2\rho_1\rho_2\cos\phi+4\left(1-\frac{2}{N}+N\right)\rho_1\rho_4\cos(\phi-\theta)\nonumber\\
&\qquad+4\left(1-\frac{2}{N}\right)\rho_1\rho_4\cos(\phi+\theta)+\frac{32}{N^2}\rho_4^2\cos2\theta+4\left(1-\frac{2}{N}+N\right)\rho_1\rho_5\cos\phi\nonumber\\
&\qquad+ 8\left(1+\frac{8}{N^2}\right)\rho_4\rho_5\cos\theta+ 4\left(1+\frac{8}{N^2}\right)\rho_4^2+ 4\left(1+\frac{4}{N^2}\right)\rho_5^2\,,\label{uni3}\\
0&=2\rho_2\rho_5+2\rho_1\rho_4\cos(\phi+\theta)-\frac{8}{N}\rho_4^2+2\left(1-\frac{4}{N}\right)\rho_4^2\cos2\theta\nonumber\\
&\qquad+2\left(3-\frac{8}{N}+N\right)\rho_4\rho_5\cos\theta-\frac{4}{N}\rho_5^2\,,\label{uni4}\\
0&=2\rho_2\rho_4\cos\theta+\left(2-\frac{8}{N}+N\right)\rho_4^2+2\left(1-\frac{4}{N}\right)\rho_4^2\cos2\theta+2\rho_1\rho_5\cos\phi \nonumber\\
&\qquad+2\left(1-\frac{8}{N}\right)\rho_4\rho_5\cos\theta+\left(2-\frac{4}{N}+N\right)\rho_5^2\,, \label{uni5}\\
0&=2\rho_1\rho_4\cos(\phi-\theta)+2\rho_2\rho_4\cos\theta+2\rho_4^2\,. \label{uni6}
\end{align}
The solutions of these equations yield the renormalization group fixed points of the $RP^{N-1}$ model in two dimensions. Since the equations were obtained relying only on the symmetries of the Hamiltonian (\ref{lattice}), their space of solutions contains the fixed points that arise in the ferromagnetic case ($J>0$) as well as in antiferromagnets\footnote{See \cite{DT1,sis} for the illustration of this point in the case of the $q$-state Potts model.} ($J<0$).

\begin{table}
\hspace{-.6cm}
{\small
\begin{tabular}{|c|c|c|c|c|c|c|c|}
\hline 
Solution & $N$ & $\rho_1$ & $\rho_2$ & $\cos\phi$ & $\rho_4$ & $\rho_5$  & $\cos\theta$ \\
\hline\hline
A1$_\pm$ & $(-\infty,\infty)$ & $0$ & $\pm 1$ & $-$ & $0$ & $0$ & $-$ \\ 
A2$_\pm$ & $[-3,2]$ & $1$ & $0$ & $\pm\frac{1}{2}\sqrt{2-M_N}$ & $0$ & $0$ & $-$\\
A3 & $-3,2$ & $\sqrt{1-\rho_2^2}$ & $[-1,1]$ & $0$ & $0$ & $0$ & $-$\\
B1 & $2$ & $\frac{1-\rho_2^2}{\sqrt{1+3\rho_2^2}}$ & $[-1,1]$ & $-\frac{2\rho_2}{\sqrt{1+3\rho_2^2}}$ & $|\rho_2|\sqrt{\frac{1-\rho_2^2}{1+3\rho_2^2}}$ & $\frac{\rho_2(1-\rho_2^2)}{1+3\rho_2^2}$ & $-\text{sgn}(\rho_2)\sqrt{\frac{1-\rho_2^2}{1+3\rho_2^2}}$\\
B$2_\pm$ & $2$ & $\sqrt{1+2x\rho_2-\rho_2^2}$ & $\alpha_\pm(x)$ & $\frac{x}{\sqrt{1+2x\rho_2-\rho_2^2}}$ & $\sqrt{\frac{-x\rho_2}{2}}$ & $\frac{-x}{2}$ & $\frac{x+2\rho_2}{2\sqrt{-2x\rho_2}}$\\
B$3_\pm$ & $3$ & $\frac{2}{3}$ & $\pm\frac{1}{3}$ & $\mp1$ & $\frac{1}{3}$ & $\pm\frac{1}{3}$ & $\pm 1$\\
\hline
\end{tabular} 
}
\caption{Analytic solutions of the equations \eqref{uni1}-\eqref{uni6}. In the expression of B$2_\pm$, $x\in\left[-\frac{1}{\sqrt{2}}\frac{1}{\sqrt{2}}\right]$ is a free parameter, and $\alpha_\pm(x)\equiv x\,\frac{2x^2-3\pm\sqrt{2(x^2-4)(2x^2-1)}}{2(6x^2+1)}$.}
\label{sol}
\end{table}

\section{Solutions}
The solutions of the equations (\ref{uni1})-(\ref{uni6}) that we could determine analytically are listed in appendix~\ref{analytic} and summarized in table~\ref{sol}. The remaining solutions, which we determined numerically, are discussed in section~\ref{numerical} below. 

Since for $\rho_4=\rho_5=0$ the equations (\ref{uni1})-(\ref{uni6}) reduce to (\ref{u1})-(\ref{u3}) with $M=M_N$, the $RP^{N-1}$ model possesses, in particular, the FPs of the $O(M_N)$ model. Notice that, for $\rho_4=\rho_5=0$, equation (\ref{sub4}) also implies $\rho_9=0$. Hence, for this class of solutions we have the vanishing of the amplidutes $S_4$, $S_5$, $S_6$ and $S_9$, namely of the amplitudes responsible for mixing indices coming from different particles (see figure~\ref{tensor_ampl}). This is why in the following we refer to these solutions as nonmixing; they are all determined analytically.  

On the other hand, not all solutions of equations (\ref{uni1})-(\ref{uni6}) are nonmixing. We now discuss the different solutions, starting from the case $N=2$. 

\subsection{$N=2$}
\begin{figure}
\centering
\includegraphics[scale=0.6]{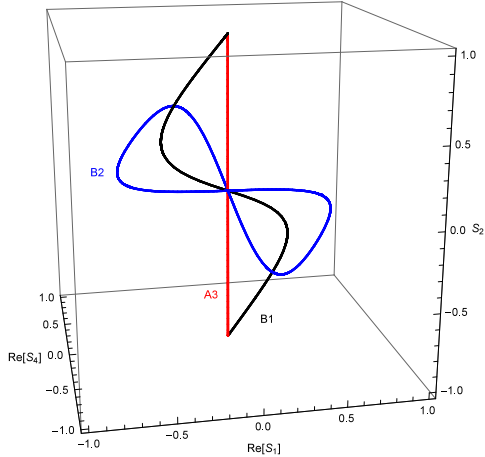}
\caption{The $N=2$ solutions A3, B1 and B2. They all meet at the BKT transition point.}
\label{plots}
\end{figure}

The equations (\ref{uni1})-(\ref{uni6}) with $N=2$ admit the solutions A3, B1 and B2 of appendix~\ref{analytic} and table~\ref{sol}. The common feature of these solutions is that they possess a free parameter, so that each of them describes a line of fixed points at $N=2$. The presence of a continuum of fixed points at $N=2$ is expected due to the topological correspondence $RP^1\sim O(2)$. The solution A3 directly corresponds to the $O(2)$ solution III of table~\ref{solutions}, which we saw accounts for the BKT transition. We now see that the $RP^1$ fixed point equations also allow for the realization of such a transition via the mixing solutions B1 and B2. This results into several lines of fixed points meeting at the BKT transition point  (figure~\ref{plots}). A similar concurrence of lines of fixed points at the BKT transition occurs in the Ashkin-Teller model, for which it was originally argued on perturbative grounds \cite{JKKN} and has recently been shown exactly \cite{DL_vector_scalar}. 

\subsection{Other solutions for $N<3$}
\label{numerical}
Besides the $N=2$ solutions of the previous subsection, the other solutions with $N<3$ of the fixed point equations \eqref{uni1}-\eqref{uni6} that we determined analytically are the solutions A1 and A2 of table~\ref{sol}. These are nonmixing solutions corresponding to I and II, respectively, of the $O(M_N)$ case (see also appendix~\ref{mapping}). The fact that $M_N$ is quadratic in $N$ is responsible for the fact that solution A3 exists also at $N=-3$ ($M_{-3}=M_2=2$), and that A2 extends down to $N=-3$. 

\begin{figure}
\centering
\includegraphics[scale=.8]{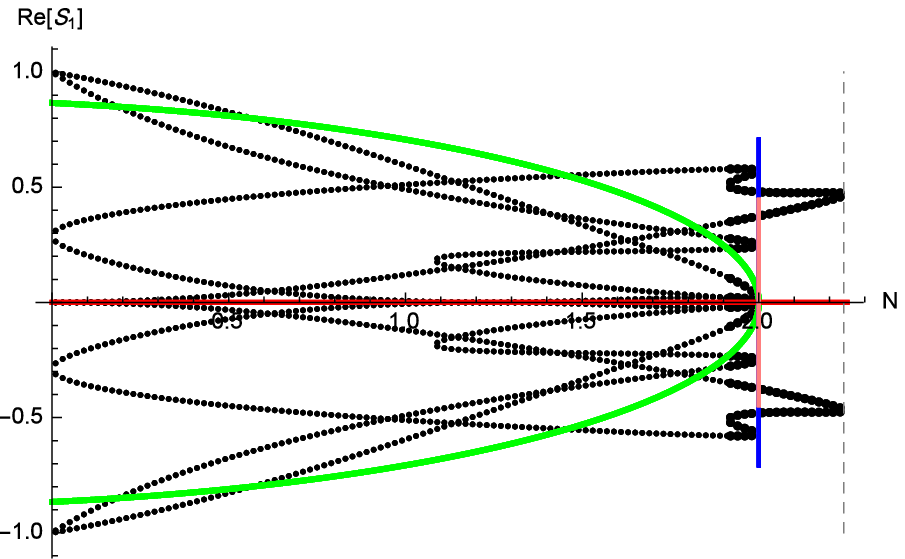}
\includegraphics[scale=.8]{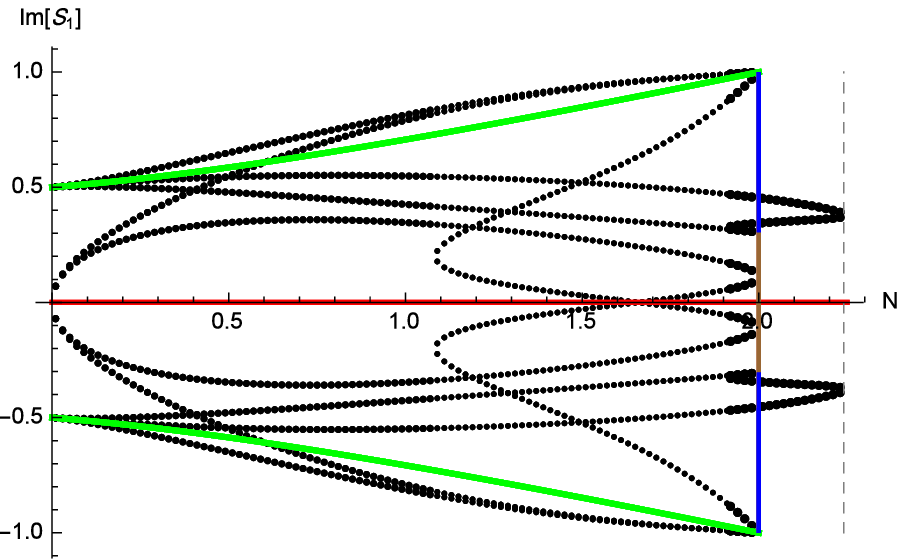}\\
\includegraphics[scale=.8]{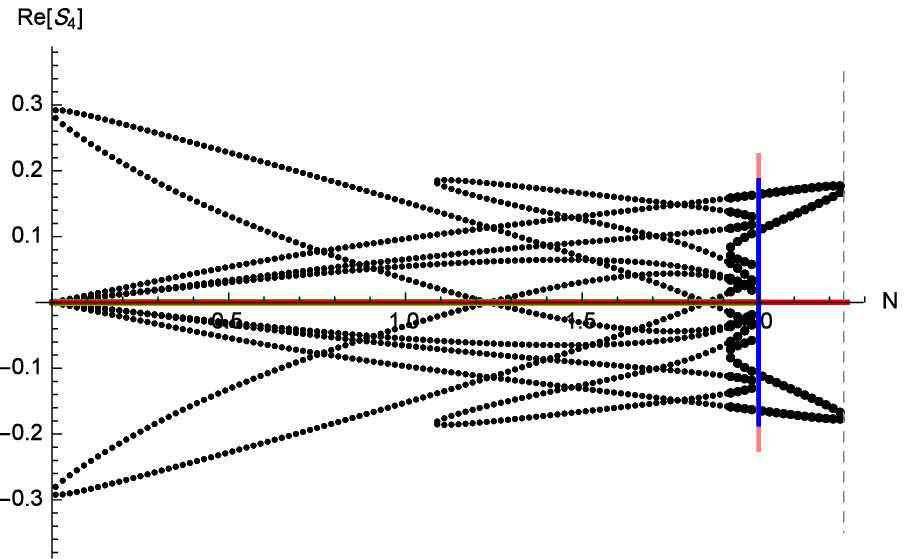}
\includegraphics[scale=.8]{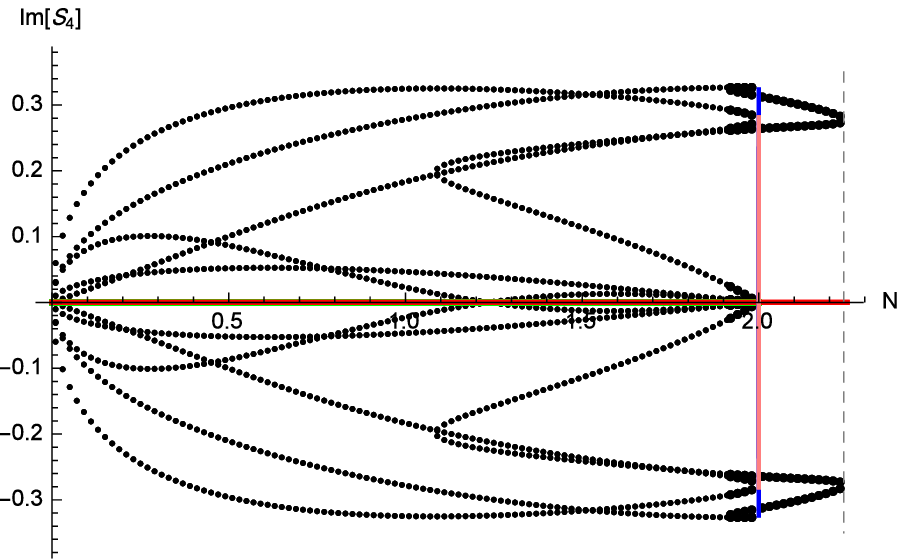}\\
\includegraphics[scale=.7]{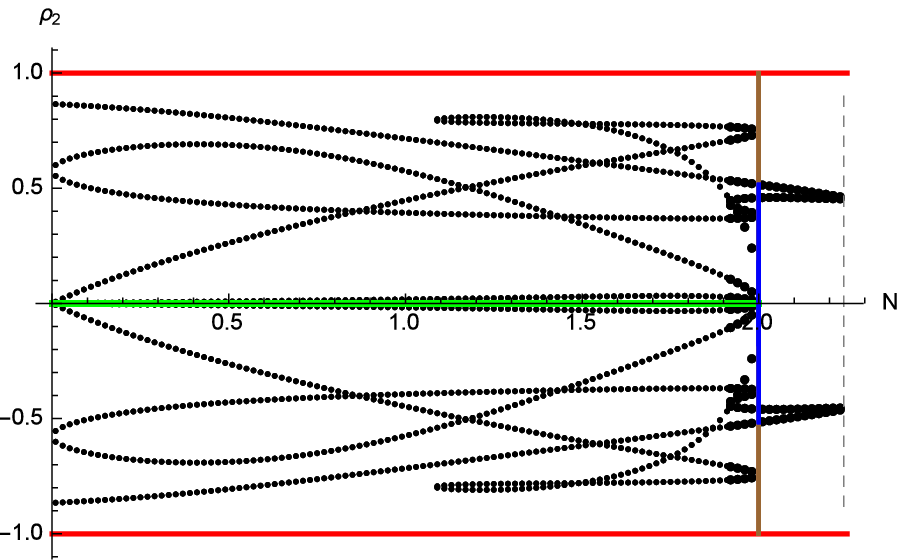}
\includegraphics[scale=.7]{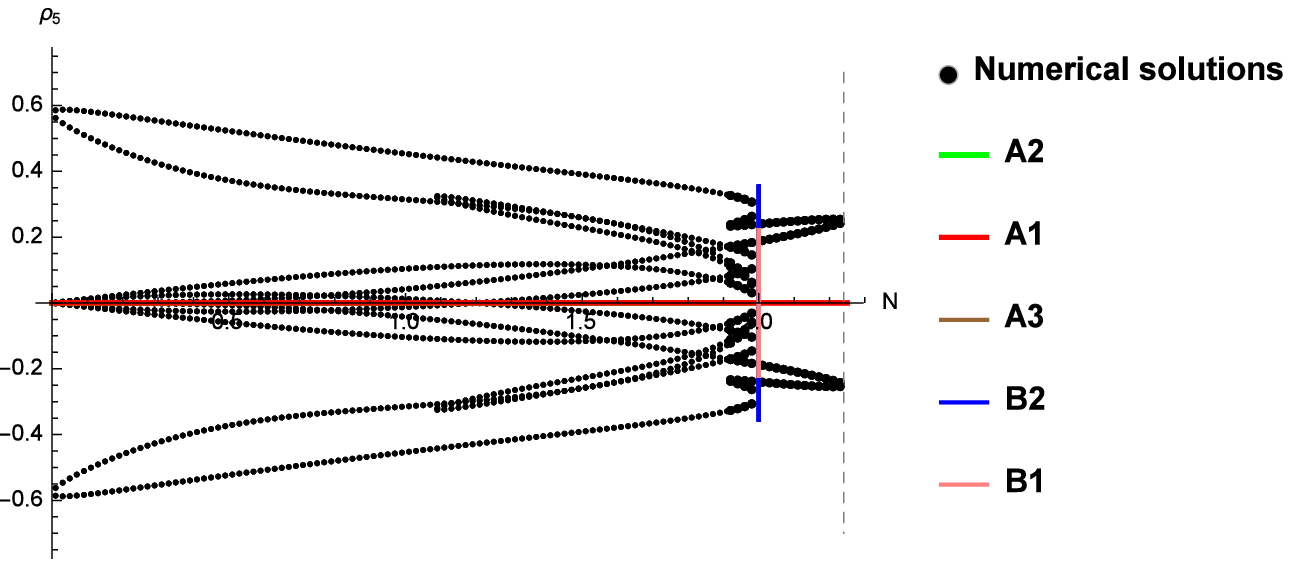}
\caption{The solutions of the fixed point equations \eqref{uni1}-\eqref{uni6} in the interval $N\in(0,N^*)$, with $N^*=2.24421..$ indicated by the dashed vertical line. The dotted lines correspond to numerical solutions, and the continuous ones to the analytic solutions of table~\ref{sol}.}
\label{numerical_solutions}
\end{figure}

Besides these analytic solutions, equations \eqref{uni1}-\eqref{uni6} admit for $N<3$ solutions that we determined numerically. All these numerical solutions are of the mixing type ($\rho_4$ and/or $\rho_5$ nonzero) and turn out to extend up to a maximal value $N^*=2.24421..$ . These solutions do not possess free parameters for fixed $N$, and for $N=2$ reduce to particular cases of the solutions discussed in the previous subsection. The numerical solutions form a rich pattern and are shown, together with the analytical ones, in figure~\ref{numerical_solutions} in the range $N\in(0,N^*)$. 

Apart from the case $N=2$, the fixed points for $N<3$ only make sense from the point of view of the analytical continuation in $N$. This is true also for $N=1$, since the $RP^0$ model possesses no degrees of freedom. On the other hand, the fixed points obtained for $N\to 1$ can have a physical meaning, in the same way that those obtained for $N\to 0$ in the $O(N)$ model are relevant for the critical behavior of self-avoiding walks.

\subsection{$N\geq 3$}
The rich pattern of fixed points with $N<3$ visible in figure~\ref{numerical_solutions} has to be contrasted with the fact that solutions A1 and B3 of table~\ref{sol} are the only fixed points existing for $N\geq 3$. The circumstance appears as a manifestation of the fact that continuous symmetries -- integer values of $N>1$ in the present case -- cannot break spontaneously in two dimensions \cite{MWHC}, thus confining criticality to zero temperature or to topological transitions. In the latter respect, the fact that the $RP^{N-1}$ fixed point solutions do not allow for free parameters at fixed $N>2$ excludes the presence in this range of BKT-like transitions yielding quasi-long-range order. The possibility of such a transition driven by disclination defects had been debated in numerical studies \cite{CPZ,KZ,FPB,DR,PFBo,FBBP,Tomita,SGR,KS}. 

Solution A1 corresponds to solution I of table~\ref{solutions}, and then we know that it yields a zero temperature fixed point in the $O(M_N)$ universality class. This is the only fixed point allowed for the $RP^{N-1}$ Hamiltonian (\ref{lattice}) for $N>3$, a conclusion that has interesting implications \cite{DDL}. In the first place, since $M_N>N$ for $N>2$, we see that the $T\to 0$ behaviors of the $RP^{N-1}$ and $O(N)$ models falls into different universality classes, thus answering a question that had been debated in the literature \cite{NWS,Hasenbusch,CHHR,CEPS,CEPS2}. In addition, the correlation length of the $O(M>2)$ model in the limit $T\to 0$ can be computed from the Hamiltonian (\ref{sigma}) and is given by \cite{Cardy_book,Zinn}
\EQ
\xi_M\propto T^{1/(M-2)} e^{A/[(M-2)T]}\,,
\label{xi}
\EN
with $A$ a positive constant. Since $\xi_M$ diverges less rapidly as $M$ increases, the identification of the $RP^{N-1}$ zero temperature critical point with the $O(M_N>N)$ critical point means that the correlation length of the $RP^{N-1}$ model diverges less rapidly than that of the $O(N)$ model, a difference that increases exponentially as $T$ decreases. This explains the very large difference in the values of the correlation length for the two models observed in numerical studies \cite{CEPS,Sinclair} and interpreted as an indication of different universality classes \cite{CEPS,CEPS2}. It also follows from our result that, for $T$ fixed, the suppression of the correlation length with respect to the $O(N)$ case decreases as $N$ increases, again in agreement with the data of \cite{CEPS} for $N=3,4$. It is worth recalling that in the $RP^{N-1}$ model the correlation length is determined by $\langle Q_{ab}(x)Q_{ab}(y)\rangle$, while $\langle{\bf s}(x)\cdot{\bf s}(y)\rangle$ vanishes as a consequence of head-tail symmetry.

Zero temperature $O(M_N)$ criticality in the $RP^{N-1}$ model corresponds to a nonmixing solution of the equations (\ref{uni1})-(\ref{uni6}), namely to $\rho_4=\rho_5=0$. Moving away from criticality, i.e. for $T>0$, these parameters are expected to acquire nonzero values, with the consequence that the detection of $O(M_N)$ behavior is likely to require very low temperatures and is numerically difficult due to the exponential divergence of the correlation length. This prediction of \cite{DDL} seems confirmed by an observation of the recent numerical study\footnote{The preprint of \cite{BFPV} appeared after that of \cite{DDL}.} \cite{BFPV}. 

These considerations are expected to apply generically for $N>2$. From this point of view, the existence for $N=3$ -- and only for this value -- of the additional solution B3 is not easy to interpret. The continuous nature of the symmetry appears to rule out a fixed point related to spontaneous breaking. On the other hand, we are not aware of a topological mechanism that could specifically apply to $N=3$. Also, a topological transition is normally associated to quasi-long-range order and to the presence of a line of fixed points, while the solution B3 corresponds to an isolated fixed point. The possibility of a zero temperature fixed point alternative to $O(M_3)=O(5)$ cannot be ruled out, but again one would like to understand why it does not arise for $N>3$. At this stage the possible role of the solution B3 remains as an interesting open question.

\section{Conclusion}
In this paper we explored the space of solutions of the exact fixed point equations of the two-dimensional $RP^{N-1}$ model for continuous values of $N\geq 0$. For $N=2$ we showed that the quasi-long-range order expected from the topological correspondence $RP^1\sim O(2)$ can actually occur in an extended form consisting of several lines of fixed points intersecting at the BKT transition point. For $N<N^*=2.24421..$ we found a rich pattern of fixed points that for $N\neq 2$ can play a role when the model is analytically continued in $N$, similarly to what happens with the loop gas description of the $O(N)$ model in the range $N\in(-2,2)$. 

The drastic rarefaction of solutions for $N>N^*$ is -- for $N$ integer -- a manifestation of the absence of spontaneous breaking of continuous symmetries in two dimensions. On the other hand, the absence for $N>2$ of lines of fixed points, i.e. of solutions containing a free parameter, shows that disclination defects do not drive a topological transition leading to quasi-long-range order at low temperatures. As a matter of fact, for generic $N>N^*$ the fixed point equations only allow for a solution corresponding to zero temperature criticality in the $O(N(N+1)/2-1)$ universality class. As observed in \cite{DDL}, this identification accounts for features of the low temperature behaviour of the correlation length observed in numerical studies. The case $N=3$ is peculiar due to the existence of an extra solution whose interpretation would require the identification of a path to criticality specific to this value of $N$. 

When comparing with numerical studies, we refer to the case of ferromagnetic interaction, since we are not aware of numerical results for $RP^{N-1}$ antiferromagnets in two dimensions. On the other hand, the fixed point equations are obtained in the scale invariant scattering framework relying only on the symmetry of the Hamiltonian. Hence, their space of solutions contains also the fixed points that can arise in antiferromagnets (this is illustrated in \cite{sis,DT1} for the $q$-state Potts model). It follows, in particular, that if zero temperature criticality is observed in a two-dimensional $RP^{N-1}$ antiferromagnet, it should fall in the $O(N(N+1)/2-1)$ universality class (with the possible caveat about $N=3$ that we pointed out above). In three dimensions the continuous symmetry can break spontaneously, and finite temperature criticality in the $O(5)$ universality class has been identified numerically for the $RP^2$ antiferromagnet on the cubic lattice \cite{Fernandez}. 

It must be observed that if, for $N\geq 2$, the square in (\ref{lattice}) is replaced by a power $p$, a first order transition is known to arise when $p$ becomes large enough \cite{ES,DSS,BGH,Vink}. Such a breakdown of universality induced by sufficiently nonlinear interactions on the lattice appears to go beyond the relation between symmetry and criticality captured by field theoretical methods. For the $RP^{N-1}$ Hamiltonian (\ref{lattice}) a first order transition was deduced for $N=\infty$ \cite{MR,KZ2} and debated for the case of $N$ large \cite{SS,TS,CMP}, while it was shown to be absent in numerical simulations performed up to $N=40$ \cite{KZ}. Our results concern the fixed points of the renormalization group, at which the correlation length diverges, and do not add to the discussion about the possibility of a first order transition at large $N$.

\appendix
\section{Analytic solutions}
\label{analytic}

We give here the analytic solutions of the fixed point equations \eqref{uni1}-\eqref{uni6}, using also \eqref{sub1}-\eqref{sub5} to express the amplitudes $S_{i>6}$.

\begin{itemize}
\item Solution A1a$_\pm$ is defined for $N \in \mathbb{R}$ and reads
\begin{equation}
\begin{split}
&\rho_2 = S_0 \,, \quad \rho_1 = \rho_4 = \rho_5 = 0,\\
& \rho_7 = \rho_8 = \rho_9 = \rho_{10} = 0.
\end{split}\label{sol1}
\end{equation}
\item Solution A1b$_\pm$ is defined for $N \in \mathbb{R}$ and reads
\begin{equation}
\begin{split}
\rho_2 &=-S_0 \,, \quad \rho_1 = \rho_4 = \rho_5 = 0, \\
\rho_7 &= \frac{2S_0}{N} \,, \quad \rho_8 = \rho_9 = 0 \,, \quad \rho_{10} = -\dfrac{\rho_7}{N}.
\end{split}\label{sol1prime}
\end{equation}
\item Solution A2$_\pm$ is defined for $N \in [-3, 2]$ and reads
\begin{equation}
\begin{split}
\rho_1 &= 1\,,\quad \cos \phi = (\pm) \dfrac{1}{2}\sqrt{2-M_N} \,, \quad \sin \phi = (\pm) \dfrac{1}{2}\sqrt{2+M_N} \,, \quad \rho_2 =  \rho_4 = \rho_5 =0, \\
\rho_7 &= \frac{S_0}{N} \,, \quad \rho_8 = \dfrac{1}{|N|} \,, \quad \psi = \pi u(N) - \phi \,, \quad \rho_9 = 0\,, \quad \rho_{10} = \dfrac{2}{N^2} \rho_1 \cos \phi - \dfrac{1}{N}\rho_7,
\end{split}
\end{equation}
with $u(N)=\begin{cases}
1\,, \hspace{0.5cm} N\geq 0\\
0\,, \hspace{0.5cm} \text{otherwise.}
\end{cases}$ 

\noindent
Here and below, signs in parenthesis are both allowed.
\item Solution A3$_\pm$ is defined for $N=-3, 2$ and reads
\begin{equation}
\begin{split}
\rho_1 &= \sqrt{1-\rho_2^2}\,, \quad \phi = (\pm) \dfrac{\pi}{2} \,, \quad \rho_2 \in [-1, 1]\,, \quad \rho_4 = \rho_5 = 0, \\
\rho_7 &= \frac{S_0}{N}-\frac{\rho_2}{N} \,, \quad \rho_8 = \frac{1}{|N|} \rho_1 \,, \quad \psi = \sgn(N) \phi\,, \quad \rho_9 = 0\,, \quad \rho_{10} = -\frac{\rho_7}{N}.
\end{split}
\end{equation}
\item Solution B$1_\pm$ is defined for $N=2$ and reads
\begin{equation} 
\begin{split}
&\rho_1=\frac{1-\rho_2^2}{\sqrt{1+3\rho_2^2}} \,, \quad \cos \phi = - \frac{2\rho_2}{\sqrt{1+3\rho_2^2}} \,, \quad \sin \phi = \pm \dfrac{\sqrt{1-\rho_2^2}}{\sqrt{1+3\rho_2^2}}\,,  \quad \rho_2 \in [-1, 1] \,,  \\[0.7em]
& \rho_4\cos\theta = -\dfrac{\rho_2(1-\rho_2^2)}{1+3\rho_2^2} \,, \quad \rho_4\sin\theta=-\frac{2\rho_2^2}{1-\rho_2^2}\rho_1\sin\phi\,,
 \quad \rho_5 = -\frac{1}{2} \rho_1\cos\phi \,, \\[0.7em]
& \rho_7=\frac{S_0}{2}-\frac{\rho_2}{2}-\rho_5 \,, \quad \rho_8=\dfrac{1}{2}\rho_1|\sin\phi| \,, \quad \psi = \pm\frac{\pi}{2}\,, \quad \rho_9=2\rho_5 \,, \quad \rho_{10}=-\frac{\rho_7}{2} \,.
\end{split}
\end{equation}
\item Solution B$2_\pm$ is defined for $N=2$ and reads
\begin{align}
& x \in \left [-\tfrac{1}{\sqrt{2}}, \tfrac{1}{\sqrt{2}} \right ] \,, \quad y = (\pm) \sqrt{1-(x - \rho_2)^2}\,, \quad \rho_2 = x\dfrac{2x^2 - 3 \pm \sqrt{2(x^2-4)(2x^2-1)} }{2(1 + 6x^2)} \nonumber \\[0.7em]
& u = \dfrac{x + 2\rho_2}{4} \,, \quad v =-\sgn(y) \sqrt{-\dfrac{x \rho_2}{2} - \left ( \dfrac{x + 2\rho_2}{4} \right )^2} \,, \quad \rho_5 = -\dfrac{x}{2} \,, \\[0.7em]
&\rho_7 =\frac{S_0}{2} + \frac{\rho_2}{2}\,, \quad p = \rho_2 + \rho_5\,, \quad  q = \dfrac{y}{2}\,, \quad \rho_9 = -2 \rho_2 \,, \quad \rho_{10} = -\dfrac{\rho_7}{2} - p \,, \nonumber
\end{align}
where $x=\rho_1 \cos \phi$, $y = \rho_1 \sin \phi$, $u = \rho_4 \cos \theta$, $v = \rho_4 \sin \theta$, $p= \rho_8 \cos \psi$, $q = \rho_8 \sin \psi$.
\item Solution B$3_\pm$ is defined for $N=3$ and reads
\begin{equation}
\begin{split}
\rho_1 &= \dfrac{2}{3} \,, \quad \phi =\pi-\theta=\pi-\psi= \dfrac{\pi}{2} \pm \dfrac{\pi}{2}  \,, \quad \rho_2 = \pm \dfrac{1}{3}\,, \quad \rho_4 = \dfrac{1}{3}\,, \quad \rho_5= \rho_2, \\
\rho_7 &=\frac{S_0}{3}+\rho_2 \,, \quad \rho_8 = \rho_1 \,, \quad \rho_9 = \mp \frac{4}{3} \,, \quad \rho_{10} = \dfrac{\rho_9-\rho_7}{3}.
\end{split}\label{sollast}
\end{equation}
\end{itemize}

The results for the scattering parameters entering equation (\ref{uni1})-(\ref{uni6}) are summarized in table~\ref{sol}. We show in the next appendix that solutions (\ref{sol1}) and (\ref{sol1prime}) only differ for the trace mode decoupling as a free boson or a free fermion; for this reason they both appear as solution A1 in table~\ref{sol}.

\section{Rewriting nonmixing solutions}
\label{mapping}
\begin{figure}
\centering
\includegraphics[scale=1.2]{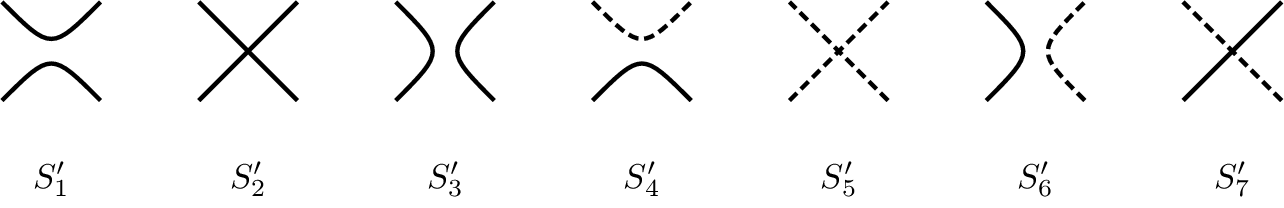}
\caption{Scattering processes of a vector particle multiplet (continuous lines) and a scalar particle (dashed lines). Time runs upwards. }
\label{vs}
\end{figure}

In this appendix we show how the nonmixing ($\rho_4=\rho_5=\rho_9=0$) solutions of the $RP^{N-1}$ fixed point equations can be written as those of a system consisting of a $O(M_N)$ vector and a scalar that are decoupled. The scattering amplitudes for such a system, in which the vector and the scalar in general interact \cite{DL_vector_scalar}, are shown in figure~\ref{vs} and take the form
\begin{align}
&S'_1=S'^*_3\equiv \rho'_1e^{i\phi'},\\
&S'_2=S'^*_2\equiv \rho'_2,\\
&S'_4=S'^*_6\equiv\rho'_4 e^{i \theta'},\\
&S'_5=S'^*_5\equiv\rho'_5,\\
&S'_7=S'^*_7\equiv \rho'_7,
\end{align}
where $\rho'_1$ and $\rho'_4$ are non negative, while $\rho'_2$, $\rho'_5$, $\rho'_7$, $\phi'$ and $\theta'$ are real. 

For the purpose of the mapping, we reorganize the particles $\mu=ab$ of the $RP^{N-1}$ model into the new basis
\begin{equation}
|\Phi_{\mu}\rangle = \begin{cases}
|{\Phi_0}\rangle=\frac{1}{N}\sum\limits_{a=1}^{N} |{aa}\rangle\\
\frac{1}{\sqrt{2}}\left(|{ab}\rangle+|{ba}\rangle\right)\,,\hspace{0.5cm} \mu=(ab), \, \, \, a \neq b \\
\frac{1}{\sqrt{k(k+1)}} \left ( \sum\limits_{j=1}^{k} |j j\rangle - k | (k+1) (k+1) \rangle \right )\,,\hspace{.5cm} \mu=kk,\, \, k = 1, \ldots N-1\, 
\end{cases}
\end{equation}
with $\langle{\Phi_\mu|\Phi_\nu}\rangle=\delta_{\mu\nu}$ and the trace mode $\Phi_0$ playing the role of the scalar of the vector-scalar system. Then, in the nonmixing case, the $RP^{N-1}$ scattering matrix can be expressed as
\begin{equation}
\begin{split}
S_{\mu, \nu}^{\rho, \sigma} &= \left (S_1^\prime \delta_{\mu, \nu} \delta^{\rho, \sigma} + S_2^\prime \delta_{\mu}^{\rho}\delta_{\nu}^{\sigma} + S_3^\prime \delta_{\mu}^{\sigma} \delta_{\nu}^{\rho} \right ) \bar{\delta}_{\mu}^0\bar{\delta}_{\nu}^0\bar{\delta}_{0}^{\rho}\bar{\delta}_{0}^{\sigma} + S_4^\prime (\delta_{\mu, \nu} \delta_{0}^\rho \delta_{0}^\sigma + \delta_{\mu}^0 \delta_\nu^0 \delta^{\rho, \sigma} ) \\ 
&\quad + S_5^\prime \delta_{\mu}^0 \delta_\nu^0 \delta_0^\rho \delta_0^\sigma + S_6^\prime (\delta_{\mu}^\sigma \delta_\nu^0 \delta_0^\rho + \delta_\mu^0 \delta_0^\sigma \delta_\nu^\rho) + S_7^\prime ( \delta_\mu^\rho \delta_\nu^0 \delta_0^\sigma + \delta_\mu^0 \delta_0^\rho \delta_\nu^\sigma),
\end{split}
\end{equation}
where $\bar{\delta}_{\mu}^{\nu} = 1- \delta_{\mu}^{\nu}$ and
\begin{align}
S_1^\prime &= \langle \Phi_\nu \Phi_\nu|\mathbb{S}| \Phi_\mu \Phi_\mu\rangle = S_1, \\
S_2^\prime &= \langle \Phi_\mu \Phi_\nu|\mathbb{S}| \Phi_\mu \Phi_\nu\rangle = S_2, \\
S_3^\prime &= \langle \Phi_\nu \Phi_\mu|\mathbb{S}| \Phi_\mu \Phi_\nu\rangle = S_3, \\
S_4^\prime &= \langle \Phi_0 \Phi_0|\mathbb{S}| \Phi_\mu \Phi_\mu\rangle = \langle \Phi_\nu \Phi_\nu|\mathbb{S}| \Phi_0 \Phi_0 \rangle = S_1 + N S_{11}, \\
S_5^\prime &= \langle \Phi_0 \Phi_0|\mathbb{S}| \Phi_0 \Phi_0 \rangle = S_1 + S_2 + S_3 + 2N(S_7 + S_8 + S_{11}) + N^2S_{10}, \\
S_6^\prime &= \langle  \Phi_\mu \Phi_0|\mathbb{S}| \Phi_0 \Phi_\mu\rangle = \langle \Phi_0 \Phi_\nu|\mathbb{S}| \Phi_\nu \Phi_0\rangle  = S_3 + NS_8, \\
S_7^\prime &= \langle \Phi_0 \Phi_\mu|\mathbb{S}| \Phi_0 \Phi_\mu\rangle = \langle \Phi_\nu \Phi_0|\mathbb{S}| \Phi_\nu \Phi_0\rangle  = S_2 + NS_7. 
\end{align}
The condition \eqref{decouplingconditions} translate into the relations
\begin{align}
& S'_4=S'_6=0 \, \, , \, \, S'_5=S'_7=S_0,
\end{align}
which express the decoupling between the vector and the scalar (see figure~\ref{vs}, recalling that $S_0=\pm 1$). The explicit form of the $RP^{N-1}$ nonmixing solutions in terms of the vector-scalar amplitudes is given in table~\ref{nm_vs}. Notice, in particular, that solutions A1a$_\pm$ and A1b$_\mp$ only differ for the fermionic or bosonic nature of the decoupled scalar.

\begin{table}[h]
\centering
\begin{tabular}{|c|c|c|c|c|c|c|c|c|}
\hline
Solution & $N$ &$\rho'_1$& $\rho'_2$ & $\cos\phi'$ & $\rho'_4$ & $\cos\theta'$ &$\rho'_5$& $\rho'_7$\\
\hline\hline
A1a$_\pm$ &$\mathbb{R}$ & $0$ & $S_0$ & $-$ & $0$ & $-$ & $S_0$ & $S_0$\\
A1b$_\pm$ &$\mathbb{R}$ & $0$ & $-S_0$ & $-$ & $0$ & $-$ & $S_0$ & $S_0$\\
A2$_\pm$  &$\left[-2,2\right]$ & $1$ & $0$ & $(\pm)\frac{1}{2}\sqrt{2-M_N}$ & $0$ & $-$ &$S_0$ & $S_0$\\
A3$_\pm$ & $2$ &$\sqrt{1-\rho_2^2}$ & $[-1,1]$ & $0$ & $0$ & $-$ & $S_0$ & $S_0$\\
\hline
\end{tabular}
\caption{Mapping between nonmixing $RP^{N-1}$ solutions and decoupled vector-scalar solutions. Signs in parenthesis are both allowed.}
\label{nm_vs}
\end{table}


\end{document}